\documentclass[epj]{svjour}

\usepackage{graphicx}
\usepackage{dcolumn}
\usepackage{bm}

\usepackage{epsfig} 
\usepackage{array} 
\usepackage{amssymb}
\usepackage{amsmath} 
\usepackage{amsbsy} 
\usepackage{epsfig}
\usepackage{mathrsfs} 
\usepackage{graphics}

\begin{document}

\newcommand{\RE}{\text{Re}}
\newcommand{\IM}{\text{Im}}
\newcommand{\eps}{\varepsilon}      
\newcommand{\bN}{{\bf N}}
\newcommand{\bM}{{\bf M}}
\newcommand{\bNn}{{\bf n}}
\newcommand{\bNm}{{\bf m}}
\newcommand{\bR}{{\bf R}}
\newcommand{\br}{{\bf r}}
\newcommand{\bX}{{\bf X}}
\newcommand{\bs}{{\bf s}}
\newcommand{\bt}{{\bf t}}
\newcommand{\cD}{{\mathscr D}}
\newcommand{\cG}{{\mathscr G}}
\newcommand{\cF}{{\cal F}}
\newcommand{\cI}{{\mathscr I}}
\newcommand{\T}{\textstyle}
\newcommand{\D}{\displaystyle \rule{0cm}{4ex}}
\newcommand{\Ds}{\displaystyle}
\newcommand{\grad}{{\boldsymbol{\nabla}}}
\newcommand{\vct}[1]{\vec{#1}}
\newcommand{\Dpartial}[1]{\frac{\partial #1}{\partial t}}
\newcommand{\DDpartial}[1]{\frac{\partial^2 #1}{\partial t^2}}
\newcommand{\Dtotal}[1]{\frac{\mathrm{d} #1}{\mathrm{d} t}}
\newcommand{\Djaumann}[1]{\frac{\mathrm{D} #1}{\mathrm{D} t}}
\newcommand{\DRjaumann}[1]{\frac{\mathrm{D}' #1}{\mathrm{D}' t}}
\newcommand{\Srate}{\dot{\boldsymbol \gamma}}
\newcommand{\srate}[1]{\dot{\gamma}_{#1}}
\newcommand{\Stress}{{\boldsymbol \tau}}
\newcommand{\stress}[1]{\tau_{\!#1}}
\newcommand{\Vort}{{\boldsymbol \omega}}
\newcommand{\vort}[1]{\omega_{#1}}
\newcommand{\RVort}{{\boldsymbol \omega}'}
\newcommand{\rvort}[1]{\omega'_{#1}}
\newcommand{\gradd}{{\boldsymbol{\nabla}}_{\!\!||}}
\newcommand{\normal}{\hat{\vct{n}}}
\newcommand{\tangtl}{\hat{\vct{t}}}
\newcommand{\pallel}{\hat{\vct{p}}}
\newcommand{\one}{{\boldsymbol 1}}
\newcommand{\trace}{\text{Tr}\,}
\newcommand{\djp}{V}
\newcommand{\pred}{p_R}


\newcommand{\be}{\begin{equation}}
\newcommand{\ee}{\end{equation}}
\newcommand{\bea}{\begin{eqnarray}}
\newcommand{\eea}{\end{eqnarray}}
\newcommand{\beas}{\begin{eqnarray*}}
\newcommand{\eeas}{\end{eqnarray*}}

\newcommand{\nn}{\nonumber}
\newcommand{\rf}[1]{(\ref{#1})}

\newcommand{\bu}{\bf u}
\newcommand{\pat}{\partial_t}
\newcommand{\pai}{\partial_i}
\newcommand{\pak}{\partial_k}
\newcommand{\pal}{\partial_l}
\newcommand{\pax}{\partial_{\!x}}
\newcommand{\pay}{\partial_{\!y}}
\newcommand{\paz}{\partial_{\!z}}
\newcommand{\pxi}{\partial_{\!\xi}}
\newcommand{\pxx}{\partial_{\!xx}}
\newcommand{\pxxx}{\partial_{\!xxx}}
\newcommand{\pxixi}{\partial_{\!\xi\xi}}
\newcommand{\pxixixi}{\partial_{\!\xi\xi\xi}}

\newcommand{\txx}{\tau^{xx}}
\newcommand{\tzz}{\tau^{zz}}
\newcommand{\txz}{\tau^{xz}}

\def\H{\mathbb{H}\,}
\def\L{\mathbb{L}\,}
\def\U{\mathbb{U}\,}
\def\W{\mathbb{W}\,}
\def\P{\mathbb{P}\,}

\title{{\it Rapid Note}\\ \\ 
A thin film model for corotational Jeffreys fluids under strong slip}

\author{Andreas M{\"u}nch\inst{1} 
\and Barbara Wagner\inst{2}
\and Markus Rauscher\inst{3,4}
\and Ralf Blossey\inst{5}}                     
\authorrunning{A. M{\"u}nch, B. Wagner}

\institute{
Institute of Mathematics, Humboldt University of Berlin, 
10099 Berlin, Germany
\and
Weierstrass Institute for Applied Analysis and Stochastics (WIAS), 
Mohrenstr. 39, 10117 Berlin, Germany
\and
Max-Planck-Institut f\"ur Metallforschung,
Heisenbergstr. 3, 70569 Stuttgart, Germany
\and
Institut f\"ur Theoretische und Angewandte Physik, 
Universit{\"a}t Stuttgart, Pfaffenwaldring 57, 70569 Stuttgart
\and
Biological Nanosystems Group,
Interdisciplinary Research Institute, c/o IEMN Avenue
Poincar\'e BP 60069, F-59652 Villeneuve d'Ascq, France
}

\date{Version: \today 
}

\abstract{
We derive a thin film model for viscoelastic liquids under strong slip 
which obey the stress tensor dynamics of corotational Jeffreys fluids. 
}


\maketitle


Dewetting liquid polymer films on nonwetting substrates,
such as silicone wafers grafted with a monolayer of brushes, 
play a prominent role in many nanotechnological applications. 
It is known that for these situations, polymer films on the 
scale of a few hundred nanometers typically show large slippage 
\cite{fetzer05}. Furthermore, for highly entangled 
polymers, the assumption of a Newtonian fluid will seize to be valid. 
To understand the interplay of viscous and viscoelastic properties of 
liquid polymers on hydrophobically coated substrates, 
there is a need for refined theoretical methods that are able to capture 
and evolve the emerging morphologies and their longtime dynamics. 
Dimension-reduced thin film models have shown in the past 
to be extremly successful to enable quantitative predictions 
that are hardly being attained simply via the underlying 
free-boundary problem. 

In this paper we make an important step in that direction by 
developing a new thin film model that combines large slippage 
with viscoelastic properties. 
In \cite{MWW05} a family of thin film 
models ranging from weak to strong slip regimes could be 
derived depending on the order of magnitude of the slip length
(see also \cite{kargupta04}).  
Modelling the viscoelastic properties of such polymers 
generalized Maxwell and Jeffreys models have been widely used. 
In \cite{RMWB05} a weak slip model could be combined with 
the linearized Jeffreys model to discuss effects of viscoelastic 
relaxation. More recently \cite{BWMR06} we have shown that
the strong slip limit can also be recovered for the linear
Jeffreys model. 
In this Rapid Note we show that, for the strong slip regime, we are 
able to fully incorporate the general corotational Jeffreys model
into our thin film model.

We begin by presenting the underlying free boundary problem 
for incompressible, viscous flow with velocity ${\bf u}$, 
\be
\grad \cdot {\bf u} = 0\, 
\ee
\be
\rho\,\Dtotal{{\bf u}} = -\grad p + \grad\cdot \Stress,
\label{eq:momentum}
\ee
where we assume that the traceless part of the symmetric
stress tensor $\Stress$ obeys the
corotational Jeffreys model \cite{BAH87}
\bea
\label{jeffreys}
\Stress + \lambda_1\frac{D\Stress}{Dt}
=
\mu\left(\Srate+\lambda_2\frac{D\Srate}{Dt}\right)\, .
\eea
Here, $D/Dt$ denotes the Jaumann derivative which for
arbitrary tensor fields $\Lambda$ is given by 
\bea
\label{jaumann}
\frac{D\Lambda}{Dt}= 
\frac{d\Lambda}{dt}+
\frac12\left({\boldsymbol\omega}\Lambda-\Lambda{\boldsymbol\omega}\right)\, ,
\eea
where $\Srate$ and $\boldsymbol\omega$ 
denote the rate of strain tensor and the vorticity tensor, given by 
\bea
\Srate=\nabla\bu+(\nabla\bu)^{\dagger},\qquad
{\boldsymbol\omega}=\nabla\bu-(\nabla\bu)^{\dagger},
\eea
respectively;
$d/dt$ is the material  derivative $\partial_t +\bu\cdot\nabla$. We assume 
the viscosity $\mu$ as well as the relaxation parameters $\lambda_1$, 
$\lambda_2$ to be constant material parameters. 

As boundary conditions we assume that the substrate is impermeable 
$\bu\cdot\vct{\hat{e}}_z=0$ at $z=0$, and further the Navier-slip 
boundary condition
\be
\bu\cdot\vct{\hat{e}}_{||} = \frac{b}{\mu}\,\Stress\cdot\vct{\hat{e}}_{||},
\ee
with a unit-vector $\vct{\hat{e}}_{||}$ parallel to the substrate (i.e., in
$x$ or $y$-direction)\/. 
At the liquid surface the normal component of the
stress is balanced by the Laplace pressure (with surface tension
coefficient $\gamma$) and the disjoining pressure $V(z)$, while the
tangential components of the stress tensor are zero. In the following we
use the reduced pressure $p_R = p+V(z)$\/.
 
To simplify the algebra we restrict the calculations to follow 
to the 2D case, where we denote the velocity field components
${\bu}=(u,w)$\/. 
We employ the strong-slip scaling, as in \cite{MWW05}. In this
limit, the friction between the liquid and the substrate is too weak to
maintain a non-zero $xz$-shear stress to lowest order, and lateral pressure
gradients are balanced by the $xx$-component of the stress tensor. For the 
stress tensor we use the same scalings as in \cite{RMWB05}, which means we 
assume that corresponding components of the strain rate and 
stress tensor are of the same order. Hence, the dimensional stress
tensor reads in terms of the non-dimensional stress tensor components
\bea
\label{2dstress}
\frac{\mu}{T}\left(
\begin{array}{cc}
\Ds\txx&\Ds\frac{\txz}{\eps}\\
\Ds\frac{\txz}{\eps}&\Ds\tzz
\end{array}
\right)
\eea
where $\mu$ denotes viscosity and $T$ denotes the time scale, 
given by $T=U/L$. $U$ denotes the 
characteristic velocity scale, such as the dewetting speed and $L$ 
the relative scale of the lateral extension of the dewetting rim. If $H$ 
denotes the relative height of the rim then we let $\eps=H/L\ll 1$. 

The dimensionless corotational Jeffreys model can then be written as
\bea
\label{eq:jeffreys1}
&&\left(1+\lambda_1\frac{d}{dt}\right)\txx-\lambda_1\left(\frac{\paz u}{\eps^2} - \pax w\right)\txz \\ \nn
&&=2\mu\left(1+\lambda_2\frac{d}{dt}\right)\pax u - \lambda_2\mu\left(\frac{(\paz u)^2}{\eps^2} - \eps^2(\pax w)^2\right)\, ,
\eea
\bea
\label{eq:jeffreys2}
&&\left(1+\lambda_1\frac{d}{dt}\right)\tzz+\lambda_1\left(\frac{\paz u}{\eps^2} - \pax w\right)\txz \\ \nn
&&=2\mu\left(1+\lambda_2\frac{d}{dt}\right)\paz w + \lambda_2\mu\left(\frac{(\paz u)^2}{\eps^2} - \eps^2(\pax w)^2\right)\, ,
\eea
\bea
\label{eq:jeffreys3}
&&\left(1+\lambda_1\frac{d}{dt}\right)\txz+\frac{\lambda_1}{2}(\paz u - \eps^2\pax w)(\txx - \tzz) \\ \nn
&&=\mu\left(1+\lambda_2\frac{d}{dt}\right)(\paz u + \eps^2\pax w)+ \lambda_2\mu\left(\paz u - \eps^2\pax w\right)\pax u\, .
\eea
These equations are coupled to the nondimensional governing equations, 
which read in the strong slip case as
\be
\pax u +\paz w =0\, ,
\ee
\bea
&&\eps^2\mbox{Re}^*\Dtotal{u}
=-\eps^2\pax p +\eps^2\pax\txx+\paz\txz\, ,
\eea
and
\bea
&&\eps^2\mbox{Re}^*\Dtotal{w}
=-\paz p +\pax\txz+\paz\tzz\, .
\eea
The corresponding scaled boundary conditions at $z=h(x,t)$ are
given by
\bea
-p_R+\frac{\eps^2\pax h^2\txx-2\pax h\txz+\tzz}{1+\eps^2(\pax h)^2}
=\frac{\pxx h}{\left(1+\eps^2(\pax h)^2\right)^{3/2}}\nn\\ \label{bc-normal}
\eea
and
\bea
\txz\left(1-\eps^2(\pax h)^2\right)-\eps^2\pax h\,\overline{\tau}=0\label{bc-tang}\, .
\eea
In the last equation we have introduced the first normal stress difference 
$ \overline{\tau} $ (commonly denoted by $N_1$ \cite{BAH87})  
\be
\overline{\tau} \equiv \txx -\tzz\, .
\ee
Finally, the kinematic condition reads
\begin{equation}
\partial_t h = - \nabla\cdot\int_0^h dz\,\, {\bf u}\, .
\end{equation} 
Note that $p_R=p + V(h)$, where $V(h)$ here denotes a disjoining pressure
due to dispersion forces across the film.

At the substrate, $z=0$, we have as usual the impermeability condition 
$w=0$ and the slip condition $u=b\txz$. In the strong-slip regime 
we have $b=\beta_s/\eps^2$, $\beta_s$ being the  
slip length parameter of order $\mathcal{O}(\eps^0)$,
see \cite{MWW05} for details. 

For the derivation of the lubrication approximation we assume that 
the dynamic variables can be given in form of the asymptotic 
expansions 
\bea
(u,w,h,p,\tau^{ij})&=&(u_0,w_0,h,p_{0},\tau^{ij}_{0})\label{ansatz}\\
&&+\eps^2 (u_1,w_1,h_1,p_{1},\tau^{ij}_1) + O(\eps^4)\, . \nn
\eea
To leading order in $\eps$ we require from equations \rf{eq:jeffreys1} 
and  \rf{eq:jeffreys2} that 
\be
\label{tau-lead}
\paz u_0=0\quad\mbox{or}\quad \txz_0=\frac{\lambda_2}{\lambda_1}\paz u_0\, .
\ee
The governing equations are, to leading order, 
\bea
&&\pax u_0+\paz w_0=0\label{mass0}\, ,\\
\nn \\
&&\paz\txz_0=0\label{momx0}\, ,\\
\nn \\
&&\paz p_{0}=\pax\txz_0+\paz\tzz_0\label{momz0}\, ,
\eea
and the leading order boundary conditions at $z=h_0$ are 
\bea
&&\txz_0=0\label{bctang0}\, ,\\
&&p_{R0}-2\left(\frac{\tzz_0}{2}-\pax h_0 \txz_0\right)+\pxx h_0=0\, ,
\label{bcnorm0}\\
&&\pat h_0-w_0+u_0\pax h_0=0\label{kin0}\, .
\eea
Leading order boundary conditions at the substrate are 
\be
w_0=0\quad\mbox{and}\quad\txz_0=0\label{bcsubstrate0}\, .
\ee
We now integrate \rf{momx0} with respect to $z$ and use the 
boundary conditions \rf{bctang0} to find
\be
\txz_0=0\label{txz0}\, .
\ee
With the constitutive equations \rf{tau-lead} and the boundary
conditions \rf{bcsubstrate0} we are led to the plug flow condition
$\paz u_0=0$. 

The next order equations for 
\rf{eq:jeffreys1}, \rf{eq:jeffreys2} are then given by
\bea
&&\left(1+\lambda_1\frac{d^0}{dt}\right)\txx_0=2\left(1+\lambda_2\frac{d^0}{dt}\right)\pax u_0\label{txx0}\, ,\\
&&\left(1+\lambda_1\frac{d^0}{dt}\right)\tzz_0=2\left(1+\lambda_2\frac{d^0}{dt}\right)\paz w_0\label{tzz0}\, ,
\eea
where ${d^0}/{dt}=\pat + u_0\pax + w_0\paz$ .
Also, from \rf{momz0}, we find with the solution \rf{txz0} and 
the boundary condition \rf{bcnorm0} for the pressure at the liquid 
surface $z=h_0$
\be
p_{0}=\tzz_0-\pxx h_0-V(h_0)\label{p0}\, .
\ee
The next order ($O(\eps^2)$) $u$-momentum equation is 
\be
\mbox{Re}^*\frac{d^0u_0}{dt}=-\pax p_{0} + \pax\txx_0 + \paz\txz_1\nn\, .
\ee
Using \rf{p0} and denoting $ u_0=f(x,t) $
we obtain
\be
\mbox{Re}^*\left(\pat f + f\pax f\right)
=
\pax \overline{\tau_0} +\pax(\pxx h_0 + V)+ \paz\txz_1\, .
\ee
Integrating this last equation from $0$ to $h_0$ we find, 
using the slip boundary conditions to the next order, 
$\txz_1=f/\beta_s$, 
\bea\label{nextorderu}
h_0 \mbox{Re}^*\left(\pat f + f\pax f\right)&=& 
\pax\int^{h_0}_0 dz\, \overline{\tau_0}\, 
- \overline{\tau_0}\vert_{z=h_0}\,\pax h_0 \\ 
& + & h_0\pax(\pxx h_0 + V) +\txz_1\vert_{z=h_0} -\frac{f}{\beta_s}\, .\nn
\eea
The next order tangential stress boundary condition at $z=h_0$ is 
\be
\txz_1=\overline{\tau_0}\,\pax h_0\label{txz1}
\ee
Hence
\bea
h_0 \mbox{Re}^*\left(\pat f + f\pax f\right)& = & \pax\int^{h_0}_0 dz\,
\overline{\tau_0} \nn \\
& + & h_0\pax(\pxx h_0+V) -\frac{f}{\beta_s}\, .
\eea
From equations \rf{txx0} and \rf{tzz0} we obtain an equation
for the difference of the diagonal terms of the stress tensor,
\bea\label{seq1}
&&(1 +  \lambda_1\pat + \lambda_1 f \pax - \lambda_1 z \pax f\paz)
\overline{\tau_0}\\
&&=4\left(\pax f +\lambda_2\pat\pax f +
\lambda_2 f(\pax f)^2 - \lambda_2 z \pax f\paz\pax f\right)\, .\nn
\eea
We now define a film-average $S$ of $\overline{\tau_0}$ as
\be
S \equiv \frac{1}{4 h_0}\int^{h_0}_0 dz\,\, \overline{\tau_0}\, . 
\label{sdef}
\ee
We denote the RHS of \rf{seq1} by $G(x,t)$ and observe that the 
last term is zero. Integrating \rf{seq1} with respect to $z$
then yields 
\bea
&&4h_0 S+\lambda_1\pat(4h_0 S)+\lambda_1 f \pax(4h_0 S)+\lambda_1(4h_0 S)\pax f\nn\\ 
&&-\lambda_1 \overline{\tau_0}\vert_{z=h_0}\left(\pat h_0 + f\pax h_0 +h_0\pax f\right) = h_0 G(x,t)\, .
\eea
Using the kinematic condition to leading order 
\be\label{kin}
\pat h_0 + \pax(f h_0) = 0
\ee
and employing the definition of $G(x,t)$ then finally leads 
to 
\be
(1 +\lambda_1\pat + \lambda_1 f \pax) S =
(\pax  + \lambda_2\pat\pax  +\lambda_2 f \pxx )f \label{seq}\, .
\ee
The lubrication model can finally be written as
\bea\label{lub1}
h_0 \mbox{Re}^*\left(\pat f + f\pax f\right) &=& \pax(4 h_0 S) 
+h_0\pax(\pxx h_0+V)  -\frac{f}{\beta_s} \nn \\
& & 
\eea
together with \rf{kin} and \rf{seq}. Note that only the advective
non-linearities, but not the corotational non-linearities appear in
\rf{seq}\/. This implies, that using the upper or lower advective
derivative instead of the Jaumann derivative (leading to Oldroyd's model A
or B) leads to the same one-dimensional thin film equation.

We are currently investigating the effect of viscoelastic 
relaxation in strongly slipping films 
on the morphology of dewetting rims.

\end{document}